\documentclass[%
 reprint,
superscriptaddress,amsmath,amssymb,
 aps,
]{revtex4-2}

\usepackage{graphicx}
\usepackage{dcolumn}
\usepackage{bm}
\usepackage{comment}
\usepackage{mathtools}
\usepackage{qcircuit} 
\usepackage{caption}
\usepackage{subcaption}
\usepackage{braket}   
\usepackage{lmodern}
\usepackage{relsize} 
\usepackage{amsmath}
\usepackage{verbatim,multirow}
\usepackage{multirow}
\usepackage[utf8]{inputenc}
\usepackage[T1]{fontenc}
\usepackage{amsmath}
\usepackage{soul}
\usepackage{amsfonts}
\usepackage{amssymb}
\usepackage{mathtools}

\newcommand{\cA}{\mathcal{A}}

\newcommand{\cE}{\mathcal{E}}
\newcommand{\cF}{\mathcal{F}}

\newcommand{\cO}{\mathcal{O}}

\newcommand{\cR}{\mathcal{R}}
\newcommand{\cS}{\mathcal{S}}

\usepackage{graphicx,color,appendix,physics,wasysym,xcolor,xparse}
\usepackage{setspace,appendix} 
\usepackage{amsmath,amssymb,amsfonts,mathtools}
\usepackage{bm,bbm,euscript,braket}
\usepackage{hyperref,cleveref}
\hypersetup{colorlinks,linkcolor={red},citecolor={blue},urlcolor={blue}}

\newtheorem{Theorem}{Theorem}
\newtheorem{lemma}{Lemma}

\newenvironment{proof}{{\bf Proof:}}{\hfill$\square$}

\usepackage{tikz}

\begin{document}

\preprint{APS/123-QED}

\title{Smallest quantum codes for amplitude-damping noise}

\author{Sourav Dutta}
\email{sourav@physics.iitm.ac.in}
\affiliation{%
Department of Physics, Indian Institute of Technology Madras, Chennai, India - 600036
}%

\author{Aditya Jain}
\email{aj722@cam.ac.uk}
\affiliation{
 University of Cambridge
}%

\author{Prabha Mandayam}
\affiliation{%
Department of Physics, Indian Institute of Technology Madras, Chennai, India - 600036
}

\begin{abstract}
We describe the smallest quantum error correcting (QEC) code to correct for amplitude-damping (AD) noise, namely, a $3$-qubit code that corrects all the single-qubit damping errors. 
We generalize this construction to a family of codes that correct AD noise up to any fixed order of the damping strength. 
We underpin the fundamental connection between the structure of our codes and the noise structure, via a relaxed form of the Knill-Laflamme conditions, different from existing formulations of approximate QEC conditions. 
Although the recovery procedure for this code is non-deterministic, our codes are optimal with respect to overheads and outperform existing codes to tackle AD noise in terms of entanglement fidelity. 
This formulation of probabilistic QEC further leads us to new family of quantum codes tailored to AD noise and also gives rise to a noise-adapted quantum Hamming bound for AD noise. 
Finally, we construct a set of universal logical gates for the $3$-qubit code, thus providing a potential pathway to fault tolerance via this class of codes.
\end{abstract}

\maketitle

\noindent \textit{Introduction.} Quantum error correction (QEC)~\cite{terhal2015} is indispensable for achieving reliable quantum computing and to scale up from the current generation of noisy intermediate-scale quantum (NISQ) devices \cite{preskill2018, Nisq01, Nisq02} to universal, fault-tolerant quantum computers. QEC involves encoding a quantum system into a proper subspace of a higher-dimensional Hilbert space. The conventional approach to QEC relies on quantum codes that are designed to correct for Pauli errors. Since the Pauli matrices form an operator basis, these codes can correct for arbitrary noise by linearity. 

However, if the noise structure of the dominant noise affecting the quantum hardware is known, one can leverage this information to construct resource-efficient quantum codes that are tailored to the noise~\cite{akshaya_review}. 
For instance, while the conventional approach requires five qubits to protect a single qubit from arbitrary single-qubit errors, there exists a four-qubit approximate quantum code tailored to amplitude-damping (AD) noise that can correct all damping errors up to single order ~\cite{Leung}. 
Subsequently, several quantum codes adapted to amplitude-damping noise have been constructed and identified, which are all of length four or higher~\cite{Fletcher, shor_2011, mg, jayashankar2020finding, dutta2024}.

It can be argued, based on the structure of the amplitude-damping channel, that the smallest quantum code to correct all single qubit amplitude-damping errors requires at least three qubits~\cite{Leung}. 
However, no known three-qubit code has yet been able to achieve this. In fact, the non-existence of a three-qubit code that satisfies the standard Knill-Laflamme conditions was proved in~\cite{non_existance} using linear programming bounds.
In this letter, we demonstrate a three-qubit code that corrects for all the single-qubit amplitude-damping errors. 
We further show that this three-qubit code satisfies a generalized form of the well-known Knill-Laflamme conditions~\cite{KLCondition}.
In a departure from previous formulations of approximate QEC~\cite{beny, prabha}, the algebraic conditions satisfied by our code allow for \emph{perfect}, syndrome-based error detection but require a \emph{non-unitary} recovery operation. 
We show how such a recovery scheme can be implemented in a probabilistic fashion with a finite success probability. 
{Our $3$-qubit QEC protocol thus falls within the framework of probabilistic or post-selected quantum error correction~\cite{nayak2006, alhejji, pQEC2, pQEC3, pQEC4}, and extends it by providing a structured, analytical recovery map.}
Finally, we show that our $3$-qubit code achieves an entanglement fidelity higher than the existing codes for single-qubit AD noise.\\
\noindent \textit{Preliminaries} Recall that a quantum channel is a completely positive trace-preserving (CPTP) map~\cite{nielsen}, whose action is described by a set of \emph{Kraus operators}. The qubit AD noise channel $\cA$ comprises of two Kraus operators~\cite{Addnoise}, labelled $A_0$ and $A_1$, which correspond to the no-damping error and the single-qubit damping error respectively, as described below.
\begin{align}
    A_0 = \ket{0}\bra{0} + \sqrt{1-\gamma} \ket{1}\bra{1}, \qquad A_1 = \sqrt{\gamma} \ket{0}\bra{1}.
    \label{eq:kraus}
\end{align}
Under the action of the AD channel, the ground state of a qubit remains unaffected, whereas the excited state decays to the ground state with probability $\gamma$. 
For multi-dimensional systems with $n$ parties, the Kraus operators take the form $A_{i_1} \otimes A_{i_2} \otimes \cdots \otimes A_{i_n} \equiv A_{i_1i_2\ldots i_n}$, where the indices $i_1,i_2,\ldots,i_n \in \{0,1\}$ for qubit systems. An error of the form $A_{i_1i_2 \ldots i_n}$ is called $t$-order error if $i_1 + i_2 + \ldots + i_n = t$. \\
\noindent \textit{A $3$-qubit Code For Amplitude-Damping Noise.} 
Consider the three-qubit code spanned by a pair of permutation-invariant states, which serve as the logical basis states,
\begin{align}\label{eq:three_qubit_code}
        \ket{0_L} = \frac{1}{\sqrt{3}} (\ket{100}+\ket{010}+\ket{001}), \ket{1_L} = \ket{111}.
\end{align}

\begin{figure}[t!]
  \centering
  \includegraphics[width=.99\linewidth]{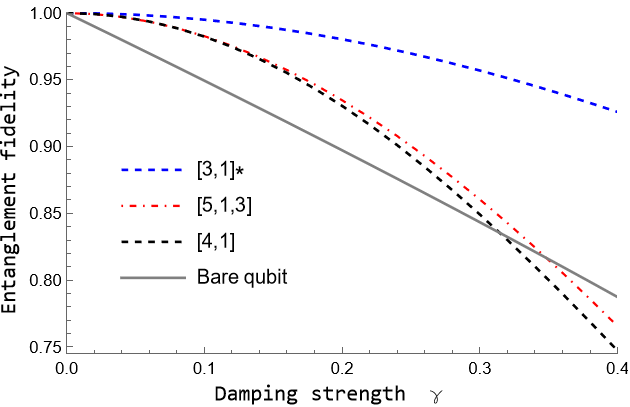}
  \caption{Entanglement fidelity $\cF_{ent}$ as a function of the damping strength $\gamma$ for our proposed $[3,1]$ (in blue dashed) AD correcting codes. The fidelities for the $[4,1]$ AD code \cite{Leung} (in red dot-dashed), the $[5,1,3]$ stabilizer code \cite{513qubit} (in black dashed), and the bare qubit (gray solid) are shown for comparison.}
  \label{fig:entfids}
\end{figure}
The no-damping error $A_{000}$ acts on the codewords as,
\begin{eqnarray}
    A_{000} \ket{0_L} &=& \sqrt{1-\gamma} \ket{0_L}, \nonumber \\
    A_{000} \ket{1_L} &=& (1-\gamma)^{\frac{3}{2}} \ket{1_L}. \label{eq:A0code}
\end{eqnarray}
The single damping errors $\{A_{100}, A_{010}, A_{001}\}$ map $\ket{0_L}$ to $\ket{000}$ up to a normalization factor of $\sqrt{\frac{\gamma}{3}}$, whereas $\ket{1_L}$ is mapped to the following states:
\begin{align}
    A_{100} \ket{1_L} = \sqrt{\gamma} (1-\gamma) \ket{011}, \nonumber\\
    A_{010} \ket{1_L} = \sqrt{\gamma} (1-\gamma) \ket{101}, \nonumber \\
    A_{001} \ket{1_L} = \sqrt{\gamma} (1-\gamma) \ket{110}. \label{eq:A1code}
\end{align}
We see from Eqs.~\eqref{eq:A0code} and~\eqref{eq:A1code} that the no-damping error and the single-damping errors map the codewords to orthogonal states. Our recovery procedure, therefore, proceeds as follows. We first perform a measurement described by the following projectors.
\begin{align*}
    P_0 &= \ket{100}\bra{100} + \ket{010}\bra{010} + \ket{001}\bra{001} + \ket{111}\bra{111}, \\
    P_1 &= \ket{000}\bra{000} + \ket{110}\bra{110} + \ket{101}\bra{101} + \ket{011}\bra{011},
\end{align*}
where $P_0$ is the projector on the codespace and corresponds to no-damping error $A_{000}$, $P_1$ is the projector on the subspace spanned by the action of the single-qubit errors on the code space.

If the measurement outcome corresponds to  $P_0$ or $P_1$, we apply appropriate recovery operators $R_0$ or $R_1$ respectively, defined as follows. The operator $R_0 = (1-\gamma) \ket{0_L} \bra{0_L}  + \ket{1_L}\bra{1_L}$ corrects for the distortion due to the no-damping error $A_{0}$. The operator $R_1 = (1-\gamma) \ket{0_L} \bra{000}  + \frac{1}{ \sqrt{3}} \ket{1_L}(\bra{110} +\bra{101} +\bra{011})$ corrects for all the single-qubit damping errors. 

One can construct a recovery channel $\cR$ using the operators $R_0P_0$ and $R_1P_1$ to get an equivalent result; however, $\cR$ is not a CPTP map in the three-qubit Hilbert space. However, we can add another ancilla qubit to construct a four-qubit CPTP recovery map $\cR_4$ with Kraus operators $\{R_0P_0 \otimes I, R_1P_1 \otimes I, \Tilde{R} \otimes X\}$, where, $\Tilde{R} = \sqrt{I - P_0 R_0^{\dagger} R_0 P_0 - P_1 R_1^{\dagger} R_1 P_1}$.
We initialize the ancilla qubit in $\ket{0}$, which gives the input state for the four-qubit recovery channel to be $\cE(\rho) \otimes \dyad{0}$. After the action of the $\cR_4$, the state becomes $(\cR \circ \cE)(\rho) \otimes \dyad{0} + \Tilde{R}\cE(\rho)\Tilde{R}^{\dagger} \otimes \dyad{1}$.
We can perform a post-selection and choose the instances only when the measurement outcome is $\ket{0}$. The post-selected state can be written as $$\Tilde{\rho} = \frac{(\cR \circ \cE)(\rho)}{\Tr[(\cR \circ \cE)(\rho)]},$$ and this can be achieved with a probability 
\begin{align}\label{eq:prob_succ}
    p_{success}= \Tr[(\cR \circ \cE)(\rho)] = (1-\gamma)^2(1-\gamma^2 \sin^2{\frac{\theta}{2}}).
\end{align}
where the input logical state is expressed as \begin{equation}\label{eq:qubit_state}
    \ket{\psi_L} = \cos{\frac{\theta}{2}} \ket{0_L} + e^{i\phi} \sin{\frac{\theta}{2}} \ket{1_L}.
\end{equation}
This means that the implementation will be successful for all input states with a probability of at least $64\%$ for a damping strength $\gamma \leq 0.2$. We discuss the implementation procedure and the details of this calculation in Section B of the supplementary material.

We benchmark the performance of our $[3,1]$ code in terms of the fidelity between the encoded and error-corrected states, expressed as $\cF_{\ket{\psi_L}} = \matrixel{\psi_L}{\Tilde{\rho}}{\psi_L}$. 
The fidelity of preserving a given input state using the $3$-qubit code is then given by,
\begin{equation}\label{eq:statefid}
    \cF_{\ket{\psi_L}} = \frac{1+\gamma^2 \sin^2{(\frac{\theta}{2})} \cos^2{(\frac{\theta}{2})} }{1+\gamma^2 \sin^2{(\frac{\theta}{2})}}.
\end{equation}
Eq.~\eqref{eq:statefid} shows that the fidelity is independent of the relative phase $\phi$.
Also, for a fixed noise strength $\gamma$, the fidelity is minimum when $\theta = \pi$, that is, for the state $\ket{1_L}$. We get the worst-case fidelity of the $[3,1]$ code given in Eq.~\eqref{eq:three_qubit_code} by substituting $\theta = \pi$ in Eq.~\eqref{eq:statefid}, that is, 
\begin{equation*}
    \cF_{\text{worst-case}} = \frac{1}{1+\gamma^2} = 1 - \gamma^2 + \cO(\gamma^3).
\end{equation*}
 The worst-case fidelity does not contain any first-order term in $\gamma$, which implies that our $[3,1]$ code can successfully correct the AD noise up to first-order in $\gamma$. In other words, the three-qubit code corrects all the single-qubit damping errors.  

We also study the performance of our code using the entanglement fidelity 
$$\cF_{ent} = \frac{\bra{\psi_p}((\cR \circ \cE)\otimes\mathbb{I})(\ket{\psi_p}\bra{\psi_p})\ket{\psi_p}}{\Tr[((\cR \circ \cE)\otimes\mathbb{I})(\ket{\psi_p}\bra{\psi_p})]},$$
where $\ket{\psi_p}$ is the purification of the logical maximally mixed state, given by $\ket{\psi_p} = \frac{1}{\sqrt{2}}(\ket{0_L}\ket{0}+\ket{1_L}\ket{1})$. 
The optimal recovery for the well known $[4,1]$ code \cite{Leung} subject to AD noise achieves an entanglement fidelity $\cF_{ent}^{[4,1]} \approx 1 - 1.25 \gamma^2 + \cO(\gamma^3)$ \cite{Fletcher_sdp} whereas our three-qubit code yields a higher entanglement fidelity $\cF_{ent}^{[3,1]}  = \frac{1}{1+ 0.5\gamma^2} = 1 - 0.5 \gamma^2 + \cO(\gamma^3)$. 
Fig. \ref{fig:entfids} compares the entanglement fidelity of our $[3,1]$ code with the $[5,1,3]$  code~\cite{513_gottesman} and the $[4,1]$ code \cite{Leung}. 
In Section E of the supplementary material, we demonstrate that the entanglement fidelity of our three-qubit code remains robust against minor experimental inaccuracies in estimating the damping strength of the AD channel. 
Finally, we note that the permutation invariant structure of the three-qubit code leads to a transversal logical $R_Z(\theta)$ gate, thus making the non-Clifford $T$ gate transversal. In fact, we obtain a universal set of logical gates for the three-qubit code, as discussed in Section H of the Supplementary material.\\

\noindent \textit{The Probabilistic QEC framework.} Although the three-qubit code defined in Eq.~\eqref{eq:three_qubit_code} does not satisfy the Knill-Laflamme conditions, it does achieve good fidelity in the presence of AD noise, by leveraging the orthogonality of the different error subspaces in the $3$-qubit space. This naturally leads to the question as to whether there exist algebraic conditions that capture the behaviour of the $[3,1]$ code in the presence of AD noise. 

\begin{figure}[t!]
  \centering
\includegraphics[width=1.0\linewidth]{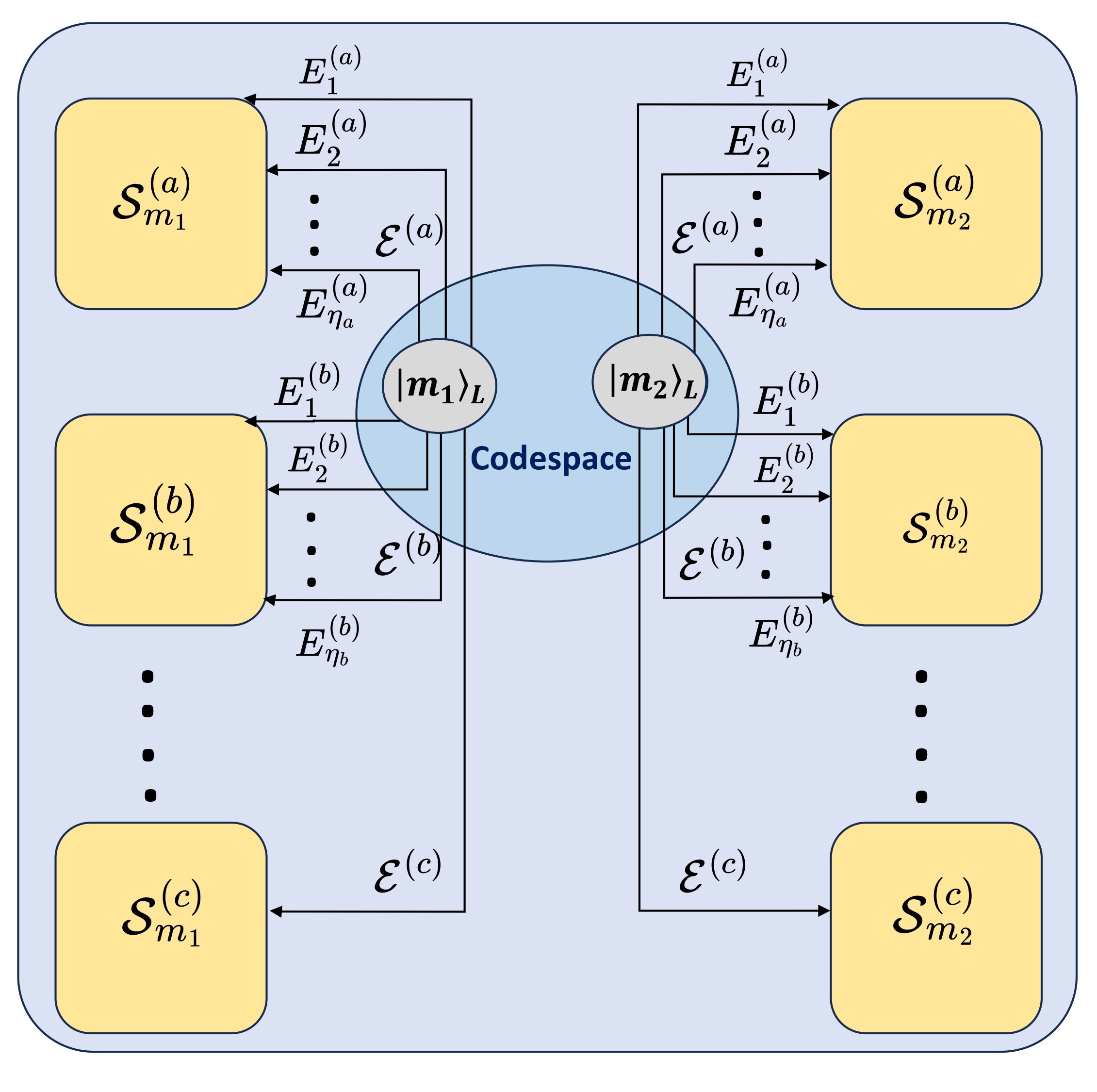}
  \caption{Representation of the action of noise on the codewords. Different sets of errors $\cE^{(a)}$ map the logical states $\ket{m_L}$ to different error subspaces $S^{(a)}_m$ which are orthogonal to each other. }
  \label{fig:fig1}
\end{figure}

Given the Kraus operators $\{E_{i}\}$ of a noise channel $\cE$, consider a grouping of the Kraus operators into sets based on their actions on the logical states. Specifically, we form sets $\cE^{(a)} = \{E_m^{(a)}\}_{m=1}^{\eta_a}$ such that the noisy states corresponding to different error sets $\cE^{(a)}$ do not overlap. In other words, for each logical state $\ket{i_L}$, the states $\{E_m^{(a)} \ket{i_L}\}_{m=1}^{\eta_a}$ (written up to appropriate normalization factors) are orthogonal to the set of states $\{E_m^{(b)} \ket{j_L}\}_{m=1}^{\eta_b}$ for all $i\neq j$ and $a \neq b$. 
Formally, we define subspaces $\cS^{(a)}_i$ spanned by the states $\{E_m^{(a)} \ket{i_L}\}_{m=1}^{\eta_a}$ and impose the constraint that these should be mutually nonoverlapping. 
Some states in the set $\{E_m^{(a)} \ket{i_L}\}_{m=1}^{\eta_a}$ can be linearly dependent, so the dimension of the subspace $\cS^{(a)}_i$ is at most $\eta_a$. 
If there are $\mu$ such groups of Kraus operators, the full Hilbert space contains $q^k \mu$ such subspaces as shown in Fig.~\ref{fig:fig1}. 
With this structure in mind, we can now state a set of sufficient conditions for probabilistic QEC, satisfied by the $3$-qubit code in Eq.~\eqref{eq:three_qubit_code}.

\begin{Theorem}\label{th:1}
Consider an $[n,k]_q$ quantum code with logical states $\{\ket{i_L}\}_{i=0}^{q^{k}-1}$ and a noise channel $\mathcal{E}$ with Kraus operators $\{E^{(a)}_m\}$, such that, 
    \begin{align}
        & \matrixel{i_L}{E^{(a)\dagger}_m E^{(b)}_{p}}{j_L} = 0, \, \forall \, m, p, \text{when~} i \neq j \text{or~} a \neq b, \nonumber\\
        & \sum_{m=1}^{\eta_a}\matrixel{i_L}{E^{(a)\dagger}_m E^{(a)}_{p}}{i_L} = \chi_{i}^{a} ~~\forall \, a, i, p ,
        \label{eq:qec_cond}
    \end{align}
where $\chi_{i}^{a}$ is a non-zero scalar depending on $i$ and $a$, then there exists a probabilistic recovery operation that perfectly corrects all the errors in the set $\{E^{(a)}_m\}$. 
\end{Theorem}

\noindent\begin{proof}
    The proof is constructive. Any logical quantum state can be expressed as a superposition of the codewords as $\ket{\psi_L} = \sum_{x} \beta_x \ket{x_L}$. The noisy state after the action of the noise channel $\cE$ is given by,
\begin{align*}
\begin{split}
    \cE(\dyad{\psi_L}) &= \sum_{w,p} E^{(w)}_p \dyad{\psi_L} E_p^{(w) \dagger} \\
    &= \sum_{p,w,x,y} \beta_x \beta^*_y E^{(w)}_p \dyad{x_L}{y_L} E_p^{(w)\dagger}
\end{split}
\end{align*}

We perform a projective measurement with the operators $\{P_1, P_2, \ldots, P_\mu, P_{\mu + 1} = \mathbb{I}-\sum_{a=1}^{\mu}P_a\}$, where $P_{a}$ is the projector onto $\cS^{(a)} \equiv \bigoplus_{i}\cS^{(a)}_{i}$, the error subspace resulting from the action of the errors in the set $\cE^{(a)}$ on the codespace. A measurement outcome corresponding to $P_a$ suggests that the quantum state is affected by one of the errors in the set $\cE^{(a)}$. We abort the protocol if we get an outcome corresponding to $P_{\mu +1}$. 

Upon detecting an error in the set $\cE^{(a)}$, the post-measurement state, up to some normalization constant, is given by,
\begin{align}
   P_a \cE(\dyad{\psi_L}) P_a = \sum_{p,x,y} \beta_x \beta^*_y E^{(a)}_p \dyad{x_L}{y_L} E_p^{(a)\dagger} . \label{eq:P_a}
\end{align}
We now define the recovery operators associated with the error set $\cE^{(a)}$ as,
\begin{align}
    R_a = \lambda_a \sum_{i=0}^{q^k -1} \frac{1}{\chi_{i}^{a}}\dyad{i_L}{i_L}\sum_{m=1}^{\eta_a}{E^{(a)\dagger}_m},
\end{align}
where, $\lambda_a$ is chosen such that the largest eigenvalue of $R_a^{\dagger}R_a$ is one. Note that the operators $\{R_aP_a\}_{a=1}^{\mu}$ forms a trace non-increasing quantum channel $\cR$. After the action of $R_a$, the state in Eq.~\eqref{eq:P_a} becomes,
\begin{align}\label{eq:recovst}
    \begin{split}
    &R_a P_a \cE(\dyad{\psi_L}) P_a R_a^{\dagger} \\
    =& \sum_{p,x,y} |\lambda_a|^2 \beta_x \beta^*_y \sum_{i,j,m,r}  \frac{1}{\chi_{i}^a(\chi_{j}^a)^*}\ket{i_L}\bra{i_L} E_{m}^{(a)\dagger} E^{(a)}_p \dyad{x_L}{y_L} \\& E_p^{(a)\dagger}E_r^{(a)}\dyad{j_L}{j_L}
    \end{split} 
\end{align}
We now use the conditions in Eq.~\eqref{eq:qec_cond} and rewrite the recovered state in Eq.~\eqref{eq:recovst} as,
\begin{equation}
    R_a P_a \cE(\dyad{\psi_L}) P_a R_a^{\dagger} = |\lambda_a|^2 \eta_a \dyad{\psi_L},
\end{equation}
where $\eta_{a}$ is the number of operators in the set $\cE^{(a)}$. The probability of successfully implementing the recovery $\cR$ is given by $\Tr\left[\cR\circ\cE(\ket{\psi_{L}}\bra{\psi_{L}})\right] = \sum_{a} |\lambda_a|^2 \eta_a$, as explained in Section B of the supplementary material. Thus we have a recovery channel $\cR$ that can correct for the noise channel $\cE$ with probability  $\sum_{a=1}^{\eta_a} |\lambda_a|^2 \eta_a$, if the codewords satisfy the conditions in Eq.~\eqref{eq:qec_cond}. 
\end{proof}

\noindent We note here a few facts related to our probabilistic quantum error correction (PQEC) conditions. 
\begin{itemize}
    \item[(i)] The probability of successfully  implementing the recovery is independent of the encoded state for the set of correctable errors, as shown above. However, when uncorrectable errors that do not satisfy the PQEC conditions are taken into account --- such as second-order and third-order damping errors for the $[3,1]$ code --- we get a state-dependent expression for the success probability, as seen in Eq.~\eqref{eq:prob_succ}.
    \item[(ii)] We can prove a weaker form of the QEC condition in Theorem~\ref{th:1}, which relaxes the requirement that the subspaces $\cS^{(a)}_i$ be mutually orthogonal. This is shown in Section G of the supplementary material.
    \item[(iii)] The $[4,1]$ code for AD noise satisfies the AQEC conditions in Eq.~\eqref{eq:qec_cond}. However, the $[3,1]$ code performs better than the $[4,1]$ code with probabilistic recovery, as shown in Section C of the supplementary material.
\end{itemize} 

\noindent\textit{A new family of quantum codes for AD noise.} We now use the PQEC conditions in Eq.~\eqref{eq:qec_cond} to construct a family of quantum codes that can correct for amplitude-damping (AD) noise using permutation-invariant quantum states. 
An $n$-qubit permutation-invariant quantum state with excitation number $e$ is defined as,
\begin{eqnarray}
   && \ket{n,e}_{PIS} = \nonumber \\
   && \frac{1}{\sqrt{\binom{n}{e}}} \sum_{\substack{x_1, x_2, \ldots ,x_n \in \{0,1\} \\
    x_1 + x_2 + \ldots + x_n = e}} \ket{x_1} \otimes \ket{x_2} \otimes \ldots \otimes \ket{x_n}. 
    \label{eq:pis}
\end{eqnarray}
We construct a family of $[2^k(t+1)-1,k]$ codes encoding $k$ logical qubits and correcting AD errors up to order $t$, whose logical states are constructed by the $n$-qubit permutation invariant states given in Eq.~\eqref{eq:pis}.
The logical state $\ket{i_L}$ is given by,
\begin{equation}
    \ket{i_L} = \ket{n,(t + 1) \text{decimal}(i) + t}_{PIS},\label{eq:codewords}
\end{equation}
where $i$ is an $n$-bit binary string, and $\text{decimal}(i)$ is the decimal number corresponding to the string $i$. This family of codes is non-additive in nature and lacks a stabilizer description.
Note that our codes differ from the permutation-invariant codes delineated in Ref.~\cite{permutation_AD}, as all the codewords of our codes have distinct excitation numbers.

The minimum number of qubits required to encode $k$ qubits using the code in Eq.~\eqref{eq:codewords} is $n_{\text{min}}=2^k(t+1)-1$, as explained in Section D.1 of the supplementary material. 
Our construction thus leads to a family of $[2^k(t+1)-1,k]$ quantum codes, which satisfy the PQEC conditions in Eq.~\eqref{eq:qec_cond}, for all the AD Kraus operators defined in \cref{eq:kraus} with damping strength up to $\cO(\gamma^t)$. We state this result formally here and refer to Section D.1 of the supplementary material for the detailed proof. 

\begin{lemma} \label{lemma:AD_qec}
    The quantum code given in Eq.~\eqref{eq:codewords} satisfies the error correction conditions in Eq.~\eqref{eq:qec_cond} for the amplitude-damping channel with damping strength up to order $t$, when the Kraus operators are grouped according to their order in terms of the damping strength $\gamma$. 
\end{lemma} 

Furthermore, for a fixed $k$, the damping strength $t$ up to which the logical qubits are protected is asymptotically linear in the total number of physical qubits $n$. Therefore, the AD analogue $(t/n)$ of relative distance $(d/n)$ is constant for this family of codes. For example, we obtain a $[5,1]$ quantum code that can correct up to second order of the damping noise as the span of the codewords,
\begin{align*}
    &\ket{0_L} \\
    &= \frac{1}{\sqrt{10}} (\ket{11000} + \ket{10100} + \ket{10010} + \ket{10001} + \ket{01100}\\&+ \ket{01010} + \ket{01001} + \ket{00110} + \ket{00101} + \ket{00011} ), \nonumber\\
    &\ket{1_L} = \ket{11111}. 
\end{align*}
A detailed comparison of the performance of our $[3,1]$ and $[5,1]$ codes against other codes in the literature is shown in Fig. 3(a) of the supplementary material. A few sample combinations of $(n,k,t)$ where a code from our family with $n$ physical qubits can protect $k$ logical qubits from AD noise up to order $t$ include $(3,1,1),(5,1,2),(7,2,1),(7,1,3),(11,2,2),(15,3,1)$.

Although the encoding rate of our code decreases exponentially with an increase in the number of logical qubits $k$ for a fixed number of physical qubits $n$ and order of correction $t$, they achieve the optimal rate possible when encoding a single qubit, that is, $k = 1$. 
We can also construct bosonic codes for the AD noise that satisfy the PQEC conditions in Eq. \eqref{eq:qec_cond} as discussed in section F of the supplementary material. 
These codes can potentially be used in photonic/optical platforms to protect against photon-loss. \\

\noindent \textit{A noise-adapted Hamming bound.}
The smallest number of qubits required to protect logical qubits from Pauli errors on a certain number of physical qubits is given by the quantum Hamming bound \cite{gottesman1997stabilizer}. Here, we obtain a noise-adapted Hamming bound, which quantifies the minimum number of qubits needed to protect the logical qubits from amplitude-damping noise of order $t$, based on the PQEC conditions in Eq.~\eqref{eq:qec_cond}.  
\begin{lemma}\label{NAhamming_saturation}
   An $[n,k]$ quantum code satisfies the PQEC conditions in Eq.~\eqref{eq:qec_cond} for amplitude-damping noise of order $t$ if and only if,
\begin{equation}\label{eq:ad_hamming_qubit}
    2^{n-k} \geq \sum_{i=0}^{t} \binom{n}{i}
\end{equation} 
The family of $[2^k(t+1)-1,k]$ qubit codes defined in Eq.~\eqref{eq:codewords} is thus optimal for $k=1$. 
\end{lemma}
A more general noise-adapted Hamming bound for the case of qudit  ($d \geq 2$) amplitude-damping noise is stated and proven in Section D.2 of the supplementary material and Eq. \eqref{eq:ad_hamming_qubit} is obtained as a special case. Proof of optimality of the family of $[2t+1, 1]$ codes can be found in Section D.3 of the supplementary material. 

\noindent \textit{Conclusions.} We have demonstrated the existence of a $3$-qubit code that can correct for first-order amplitude-damping noise, by going beyond the current framework of approximate quantum error correction. This code works by grouping the set of correctable errors in such a way that distinct error subsets can be distinguished by unique projective measurements. The non-unitary action of the errors makes the recovery protocol probabilistic, but we show that this protocol can be implemented with a finite probability of success. Notably, our $3$-qubit code has now been successfully implemented in an IBMQ processor with break-even performance, thus marking the first demonstration of noise-adapted quantum error correction on a publicly accessible quantum hardware~\cite{Vismay_2026}.

Our generalized PQEC conditions lead to a family of quantum codes that encode logical states into permutation invariant states with different excitation numbers. These codes exhibit superior performance against AD noise compared to all existing quantum codes in terms of entanglement fidelity. Our approach also enables us to write down a noise-adapted quantum Hamming bound that is tailored for AD noise. 

The PQEC framework presented here can be used to find efficient quantum codes when the dominant noise process of the hardware is known and has a non-unitary structure. An immediate line of investigation would be to find efficient noise-adapted quantum codes for other physically motivated non-unitary noise processes, such as photon loss and generalized AD noise. Unlike other noise-adapted quantum codes for AD noise~\cite{Leung, dutta2024}, the codes presented here do not have any stabilizer structure, and hence fall under the family of non-additive codes. 

Investigating fault tolerance aspects of such families of codes is another interesting direction for future research. Our construction of a universal set of logical gates for the $3$-qubit AD code is a first step in this direction. The fact that these are non-stabilizer, non-additive codes leads to novel and interesting constructions for the logical gates, as exemplified by the transversal $T$ gate. Investigating whether these logical gates can lead to fault-tolerant gadgets and evaluating a threshold along the lines of~\cite{Aj, my2025fault} is left for future work.

\noindent\textit{Acknowledgements.} We thank Debjyoti Biswas for useful discussions and Markus Grassl for insightful comments on an earlier version of this draft. This research was supported in part by a grant from the Mphasis F1 Foundation to the Centre for Quantum Information, Communication, and Computing (CQuICC). AJ acknowledges support from ERC Starting Grant 101163189 and UKRI Future Leaders Fellowship MR/X023583/1. We are grateful to the anonymous reviewer for their constructive suggestions.

\bibliography{apssamp}

\end{document}